\documentclass[12pt]{article}
\newlength{\extraspace}
\newlength{\extraspaces}
\catcode`\@=11
\def\numberbysection{\@addtoreset{equation}{section}
\def\theequation{\arabic{section}.\arabic{equation}}}
\newcommand{\newsection}[1]{
\vspace{7mm}
\pagebreak[3]
\addtocounter{section}{1}
\setcounter{subsection}{0}
\setcounter{footnote}{0}
\begin{center}
{\large {\bf \thesection. #1}}
\end{center}
\nopagebreak
\medskip
\nopagebreak
\hspace{3mm}}
\newcommand{\nonu}{\nonumber \\[.5mm]}
\newcommand{\A}{&\!\!\!}

\newcommand{\be}{\begin{equation}}
\newcommand{\bea}{\begin{eqnarray}}
\newcommand{\eea}{\end{eqnarray}}
\newcommand{\ba}{\begin{array}}
\newcommand{\ea}{\end{array}}
\newcommand{\ee}{\end{equation}}

\newcommand{\br}{\begin{array}}
\newcommand{\er}{\end{array}}
\begin{document}
\addtolength{\baselineskip}{.7mm}
\thispagestyle{empty}
\begin{flushright}
MIT--CTP--3381 \\
{\tt hep-th/0306095} \\ 
June, 2003
\end{flushright}
\vspace{7mm}
\begin{center}
{\Large{\bf A Note on Supersymmetry 
in Noncommutative Field Theories 
}} \\[20mm] 
{\sc Madoka Nishimura}\footnote{
\tt e-mail: madoka@mit.edu} 
\hspace{1mm}  \hspace{2mm}
\\[7mm]
{\it Center for Theoretical Physics \\
Massachusetts Institute of Technology \\
Cambridge MA 02139-4307, USA} \\[20mm]
{\bf Abstract}\\[7mm]
{\parbox{13cm}{\hspace{5mm}
We show that a solution of the type IIB supergravity representing 
D3-branes in the presence of a 2-form background has 16 
supersymmetries by explicitly constructing the transformation 
parameter of unbroken supersymmetry. 
This solution is dual to a noncommutative Yang-Mills theory 
in a certain limit. 
}}
\end{center}
\vfill
\newpage
\setcounter{section}{0}
\setcounter{equation}{0}
%
\newsection{Introduction}
%
The local supersymmetry in supergravities is directly 
connected to other local symmetries, such as the gauge symmetry 
and the general covariance as well as the local Lorentz 
symmetry, as can be seen in the commutator algebra of the 
transformations. 
On the other hand, noncommutative theories can have supersymmetry 
while the Lorentz symmetry is broken by the presence of a 
constant 2-form background $B_{\mu\nu}$. It is conjectured 
that a noncommutative super Yang-Mills theory 
with constant $B_{23}$ has 16 supersymmetries \cite{SW,HI}. 
Thus, it is interesting to see the relation between the 
supersymmetry and the local Lorentz symmetry in the 
supergravity, which is dual to the noncommutative field theory 
with the Lorentz symmetry intuitively broken. 
\par
In this paper we show that a solution of the type IIB supergravity 
representing D3-branes in the presence of non-vanishing 
$B_{01}$, $B_{23}$ backgrounds has 16 supersymmetries, although 
the Lorentz symmetry is broken on the world volume. 
This solution was obtained in ref.\ \cite{MR} by a Wick rotation 
from the Euclidean solution representing smeared bound states 
of D($-1$), F1, D1 and D3 branes. 
It originally appeared in the context of the T-duality 
map \cite{BRGPT}, and was discussed as D-brane bound states 
\cite{RT,BMM}. 
In order to study supersymmetry 
we use the Lorentzian solution rather than the Euclidean one since 
it is not clear how to define supersymmetry in the Euclidean space. 
\par
We substitute this solution into the supertransformations of the 
fermionic fields in the type IIB supergravity and obtain the 
conditions on the transformation parameter of unbroken supersymmetry.  
The general form of the parameter satisfying these conditions contains 
an arbitrary constant spinor which has 16 independent components. 
Therefore, we find that 16 supersymmetries exist in four dimensions. 
This remains true even when we turn off $B_{01}$ or when we take 
the decoupling limit \cite{MR}. 
We will see that the parameter of unbroken supersymmetries 
explicitly depends on the scalar field $\tau$, the angles of 
the D-branes tilted in the 01 and 23 directions and the harmonic 
functions appearing in the metric. 
\par
The number of supersymmetries may be partly explained by 
the isometry SO(6) = SU(4) of S${}^5$ contained in the solution. 
The solution with vanishing 2-form background has ${\cal N}=4$ 
Poincar\'e supersymmetry in four dimensions \cite{MAL}, 
whose four-component spinor transformation parameters are 
independent of the four-dimensional coordinates $x^\mu$. 
It also has the conformal supersymmetry with $x^\mu$-dependent 
transformation parameters. 
We will see that the solution with non-vanishing 2-form 
background has the same number of supersymmetries with 
$x^\mu$-independent parameters, 
but the spinor transformation parameters satisfy a certain 
condition and have only two independent components. 
There is no conformal supersymmetry in this case. 
\par
In refs.\ \cite{ST,ST2} the D-brane effective action was 
obtained as a solution to the Hamilton-Jacobi equation for the 
type IIB supergravity compactified on S${}^5$. 
The ADM formalism for the 5-dimensional space-time was used 
by treating the radial coordinate $r$ as time. 
Although only the bosonic part of the effective action has been 
studied, it brought about remarkable results to 
find new supergravity solutions. 
The effective action obtained there includes a broad family of 
super Yang-Mills theories by only assuming that 2-form fields are 
constant at the surface of constant radius $r$. 
The geometries dual to the field theories described by this 
effective action are not known in general. However, there are  
some known geometries which are dual to 
the field theories brought by the Hamilton-Jacobi formalism. 
The noncommutative Yang-Mills theory and the ordinary super Yang-Mills 
theory are solutions to the Hamilton-Jacobi equation. 
The dual geometry of the noncommutative Yang-Mills theory is 
a near horizon limit of a supergravity solution of $N$ D3-branes 
in a non-vanishing $B_{23}$ background. 
The dual geometry of the ordinary super Yang-Mills theory is 
obtained in the limit of vanishing $B_{23}$ \cite{IMSY}. 
\par
To know more about the general super Yang-Mills theories which is 
derived as a solution to the Hamilton-Jacobi equation, 
and to look for the corresponding dual geometries, 
it is useful to consider the solution with a non-vanishing 
2-form extended in the 01 and 23 directions and non-vanishing 0-forms, 
and to see how supersymmetries survive. 
It is also important to look at the correspondence between the 
symmetries of the supergravity and those of the field theory  
in the light of the AdS/CFT correspondence, as the maximal
supersymmetric case shows that the local symmetries in 
the supergravity become the global symmetries of the field theory 
on the boundary \cite{NT}. 
\par
We write down the supertransformations of the type IIB supergravity 
in the string frame in Section 2, and give the explicit form of the 
solution with non-vanishing 2-form fields in Section 3. 
In Section 4 we substitute the solution into the 
supertransformations and find the parameter of unbroken supersymmetry. 
\par
%
\newsection{Type IIB supergravity in ten dimensions}
%
The field content of the type IIB supergravity in ten 
dimensions \cite{SCHWARZ,HW} is a vielbein $e_M{}^A$, 
a complex Rarita-Schwinger field $\psi_M$, 
a real fourth-rank antisymmetric tensor field $D_{MNPQ}$ with a 
self-dual field strength, two real second-rank antisymmetric 
tensor fields $B_{MN}$, $C_{MN}$, 
a complex spinor field $\lambda$ and a complex scalar field 
$\tau = \chi + i e^{-\phi}$. 
We use the string frame rather than the Einstein frame. 
In the string terminology $\chi$, $C_{MN}$ and $D_{MNPQ}$ are 
0, 2, 4-form fields in the R-R sector, while $B_{MN}$ and 
$\phi$ are a 2-form field and a dilaton in the NS-NS sector. 
We denote ten-dimensional world 
indices as $M,N,\cdots = 0,1,\cdots,9$ and local Lorentz indices 
as $A,B, \cdots = 0,1,\cdots, 9$. 
The fermionic fields satisfy chirality conditions 
$\bar\Gamma_{10D} \psi_M = - \psi_M$, 
$\bar\Gamma_{10D} \lambda = \lambda$, 
where $\bar\Gamma_{10D} = \Gamma^0 \Gamma^1 \cdots \Gamma^9$ 
is the ten-dimensional chirality matrix. 
We choose the ten-dimensional gamma matrices $\Gamma^A$ to 
have real components. 
The gamma matrices with the world indices are denoted as 
$\hat{\Gamma}^M = \Gamma^A e_A{}^M$. 
\par
The local supersymmetry transformations of the fermionic fields 
in the string frame can be obtained from those in the Einstein 
frame \cite{SCHWARZ,HW} and are given by 
\begin{eqnarray}
\kappa \delta \lambda 
\A = \A - e^{{1 \over 4}\phi} \left( \hat{\Gamma}^M \epsilon^* P_M 
+ {1 \over 24} e^\phi \hat{\Gamma}^{MNP} 
\epsilon {\cal G}_{MNP}\right), \nonu
\kappa \delta \psi_M 
\A = \A \left[ 
	 \partial_M 
	 +{1\over 4}\left(\omega_M{}^{AB}\Gamma_{AB}
	 +{1\over 2}\partial_N \phi\hat{\Gamma}^N{}_M \right)
	 -{1 \over 2}i Q_M \right]\epsilon \nonu
\A\A + {1 \over 16\cdot 5!} i e^\phi \tilde{F}_{P_1 \cdots P_5} 
\hat{\Gamma}^{P_1 \cdots P_5} \hat{\Gamma}_M \epsilon 
\nonu
\A\A - {1 \over 96} e^\phi \left( \hat{\Gamma}_M{}^{NPQ} {\cal G}_{NPQ} 
- 9 \hat{\Gamma}^{PQ} {\cal G}_{MPQ} \right) \epsilon^*, 
\label{susytrans}
\end{eqnarray}
where $\kappa$ is the gravitational coupling constant and 
\begin{eqnarray}
P_M \A = \A {1 \over 2} i e^\phi 
{1+i \tau^* \over 1 - i \tau} \partial_M \tau, \nonu
Q_M \A = \A - {1 \over 4} e^\phi 
 \left( {1-i\tau^* \over 1-i\tau} \partial_M \tau 
  + {1+i\tau \over 1+i\tau^*} \partial_M \tau^* \right), \nonu
{\cal G}_{MNP} \A = \A i \left({1+i\tau^* \over 1-i\tau}\right)^{1 \over 2}
\left( F_{MNP} + \tau H_{MNP} \right), \nonu
\tilde{F}_{MNPQR} \A = \A F_{MNPQR} + 10 C_{[MN} H_{PQR]}. 
\end{eqnarray}
The field strengths are defined as 
\begin{eqnarray}
F_{MNPQR} \A = \A 5 \partial_{[M} D_{NPQR]}, \quad 
F_{MNP} = 3 \partial_{[M}C_{NP]}, \quad 
H_{MNP} = 3 \partial_{[M} B_{NP]}. 
\end{eqnarray}
The 5-form field strength $\tilde{F}_{MNPQR}$ must satisfy 
the self-duality condition 
$*\tilde{F}_{M_1 \cdots M_5}=\tilde{F}_{M_1\cdots M_5}$. 
The transformation parameter $\epsilon$ is a complex spinor 
satisfying the chirality condition 
$\bar\Gamma_{10D} \epsilon = - \epsilon$. 
The unusual term proportional to $\partial_N \phi$ in 
$\delta\psi_M$ is due to the use of the string frame. 
\par
For later convenience, we redefine the transformation 
parameter as 
\begin{equation}
\tilde{\epsilon} 
= \left({1+i\tau^* \over 1-i\tau}\right)^{-{1 \over 4}}\epsilon, 
\qquad
\tilde{\epsilon}^* 
= \left({1+i\tau^* \over 1-i\tau}\right)^{1 \over 4}\epsilon^*.
\end{equation}
It is easy to show that in terms of $\tilde{\epsilon}$ the 
derivative term in $\delta\psi_M$ looks simpler 
\begin{equation}
 \left( \partial_M -{1\over 2}iQ_M \right) \epsilon
= \left({1+i\tau^* \over 1-i\tau}\right)^{1 \over 4}
\left( \partial_M + {1 \over 4} i e^\phi \partial_M \chi \right)
\tilde{\epsilon}.
\end{equation}
\par
%
\newsection{The solution with non-vanishing 2-form fields}
%
The noncommutative super Yang-Mills theory is described by 
a solution of the type IIB supergravity with non-vanishing $B_{MN}$ 
field \cite{MR} in the Seiberg-Witten limit \cite{SW}. 
Before taking the limit the solution has the metric 
\begin{eqnarray}
ds_{\rm string}^2 \A = \A G_{MN} dx^M dx^N \nonu
\A = \A f^{-1/2} h' \left[ -(dx^0)^2 + (dx^1)^2 \right] 
+ f^{-1/2} h \left[ (dx^2)^2 + (dx^3)^2 \right] \nonu
\A\A +f^{1/2} \delta_{ij}dx^idx^j, 
\label{metric}
\end{eqnarray}
where 
\begin{eqnarray}
f \A = \A 1+{\alpha'^4 R^4 \over r^4}, \qquad
r^2 = \delta_{ij} x^i x^j, \nonu
h^{-1} \A = \A \sin^2 \theta f^{-1} + \cos^2 \theta, \qquad 
h'^{-1} = - \sinh^2 \theta' f^{-1} + \cosh^2 \theta'. 
\end{eqnarray}
The radius $R$ is given by 
$R^4 = 4\pi gN (\cos \theta \cosh \theta')^{-1}$. 
We have decomposed the world indices as $M=(\mu,i)$ 
($\mu=0,1,2,3$; $i=4,5, \cdots,9$). 
Other non-vanishing fields are 
\begin{eqnarray}
e^{2\phi} \A = \A g^2 hh', \qquad 
\chi = -g^{-1} \sin \theta \sinh \theta' f^{-1}, \nonu
B_{01} \A = \A \tanh \theta' f^{-1} h', \qquad
B_{23} = \tan \theta f^{-1} h, \nonu
C_{01} \A = \A g^{-1} \sin \theta \cosh \theta' h' f^{-1}, \qquad
C_{23} = - g^{-1} \cos \theta \sinh \theta' h f^{-1}, \nonu
D_{0123} \A = \A g^{-1} \cos \theta \cosh \theta' hh' f^{-1}  
\label{solutions}
\end{eqnarray}
and $D_{ijkl}$ determined from the above $D_{0123}$ by the 
self-duality condition of $\tilde{F}_{MNPQR}$. 
The constants $\theta$ and $\theta'$ parametrize the asymptotic 
values of $B_{23}$ and $B_{01}$ for $r \rightarrow \infty$. 
The existence of non-vanishing $\chi$, $B_{01}$ and $C_{01}$ 
means that D($-1$), F1 and D1 branes enter into the bound state. 
\par
We also need the spin connection $\omega_M{}^{AB}$ for the metric 
(\ref{metric}). From the torsionless condition 
$\partial_{[M} e_{N]}{}^{A} + \omega_{[M}{}^{AB} e_{N]B}= 0$
the non-vanishing components are obtained as 
\begin{eqnarray}
\omega_\mu{}^{aI} \A = \A 
-{1\over 4} \left( 1 + 2 \sinh^2 \theta' f^{-1} h' \right) 
{h'}^{1 \over 2} f^{-{3 \over 2}} \partial_i f \delta^{iI} \delta_\mu^a 
\qquad (\mbox{for }\mu=0,1), \nonu
\omega_\mu{}^{aI} \A = \A 
-{1\over 4} \left( 1 - 2 \sin^2 \theta f^{-1} h \right) 
h^{1 \over 2} f^{-{3 \over 2}} \partial_i f \delta^{iI} \delta_\mu^a 
\qquad (\mbox{for }\mu=2,3), \nonu
\omega_i{}^{IJ} \A = \A 
-{1\over 2}f^{-1}\partial_j f \delta^{j[I}\delta_i^{J]}, 
\end{eqnarray}
where we have decomposed the local Lorentz indices as $A=(a, I)$ 
($a = 0,1,2,3$; $I = 4,5,\cdots,9$). 
For later convenience, we define 
$\hat\Gamma^r = {x^i \over r} \hat\Gamma^i$ 
as the gamma matrix for the radial ($r$) direction. 
\par
%
\newsection{Supersymmetry with non-vanishing 2-form fields}
%
The final task is to substitute the solution (\ref{metric}), 
(\ref{solutions}) into the supersymmetry transformations 
(\ref{susytrans}) and obtain the conditions on the transformation 
parameter $\epsilon$ for unbroken supersymmetry. 
Since the fields in eqs.\ (\ref{metric}), (\ref{solutions}) depend 
only on the radial coordinate $r$, 
some expressions in the supertransformations become simpler. 
For example, we obtain  
\begin{eqnarray}
{\cal G}_{\mu\nu r} 
= i\left( {1+i\tau^* \over 1-i\tau} \right)^{1 \over 2}
\left(\partial_rC_{\mu\nu}+\tau \partial_rB_{\mu\nu}\right) 
\end{eqnarray}
and other components of ${\cal G}_{MNP}$ vanish. 
Only the $M=r$ components of $P_M$, $Q_M$ are non-vanishing. 
The vanishing of the supertransformation of $\lambda$ in 
eq.\ (\ref{susytrans}) then requires 
\begin{equation}
\epsilon^* = - {1\over 8} e^{\phi} {\cal G}_{\mu\nu r} 
\hat{\Gamma}^{\mu\nu}\epsilon \left(P_r\right)^{-1}. 
\label{killing}
\end{equation}
Substituting (\ref{solutions}) we obtain more explicit condition 
\begin{equation}
\tilde{\epsilon}^*
= - f^{1 \over 2} \left(\sin\theta h^{1 \over 2} 
+ i\sinh\theta' {h'}^{1 \over 2} \right)^{-1}
\left( h^{1 \over 2} \cos\theta \Gamma^{0123} 
- i{h'}^{1 \over 2} \cosh\theta'\right)
\Gamma^{01} \tilde{\epsilon}. 
\label{killing2}
\end{equation}
\par
%
We then consider the condition from $\delta\psi_M = 0$. 
By using the self-duality of $\tilde{F}_{M_1 \cdots M_5}$ and 
the chirality condition of $\epsilon$ we find that 
the 5-form field strength term in $\delta\psi_M$ becomes 
\begin{equation}
{1 \over 5!} \tilde{F}_{NPQRS} \hat{\Gamma}^{NPQRS} 
\hat{\Gamma}_M \tilde{\epsilon}
= 2 \tilde{F}_{0123r} \hat{\Gamma}^{0123} \hat{\Gamma}^r 
\hat{\Gamma}_M \tilde{\epsilon}, 
\label{5form}
\end{equation}
where $\tilde{F}_{0123r} = \partial_rD_{0123}+C_{01}\partial_rB_{23}
+C_{23}\partial_rB_{01}$. 
The ${\cal G}$ terms are simplified as 
\begin{equation}
-{e^\phi\over 96}\left(3 \hat{\Gamma}_\mu{}^{\nu\rho i}
{\cal G}_{\nu\rho i}
-18 \hat{\Gamma}^{\nu i}{\cal G}_{\mu\nu i}\right)
=-{e^\phi\over 16} f^{1 \over 2} 
\left( h^{-1} {\cal G}_{23r} \Gamma^{0123}
+ 3 {h'}^{-1} {\cal G}_{01r}\right) 
\Gamma^{01}\hat{\Gamma}_\mu{}^r 
\label{3form1}
\end{equation}
for $\mu=0,1$, 
\begin{equation}
-{e^\phi\over 96}\left(3 \hat{\Gamma}_\mu{}^{\nu\rho i}
{\cal G}_{\nu\rho i}
- 18 \hat{\Gamma}^{\nu i}{\cal G}_{\mu\nu i}\right)
=-{e^\phi\over 16} f^{1 \over 2} 
\left( {h'}^{-1} {\cal G}_{01r}
+ 3 h^{-1} {\cal G}_{23r} \Gamma^{0123} \right) 
\Gamma^{01}\hat{\Gamma}_\mu{}^r 
\label{3form2}
\end{equation}
for $\mu=2,3$, and 
\begin{equation}
-{e^\phi\over 96}\left(3 \hat{\Gamma}_i{}^{\nu\rho j}
{\cal G}_{\nu\rho j}
-9 \hat{\Gamma}^{\mu\nu}{\cal G}_{\mu\nu i}\right)
= - {e^\phi\over 16}{x^j\over r} f^{1 \over 2} 
\left( {h'}^{-1} {\cal G}_{01r} 
+ h^{-1} {\cal G}_{23r} \Gamma^{0123} \right) \Gamma^{01}
\left( \hat{\Gamma}_i{}^j - 3\delta_i^j \right)
\label{3form3}
\end{equation} 
for $i=4, \cdots, 9$. Note that the summations of $\nu$, $\rho$ 
on the left-hand sides run 0 to 3. 
\par
We use these formulae and substitute eq.\ (\ref{killing}) into 
$\epsilon^*$ in $\delta\psi_M$. 
For $\mu=0,\cdots,3$ the spin connection term and the 
$\partial_j \phi$ term cancel with the real part of the 3-form 
terms in eqs.\ (\ref{3form1}), (\ref{3form2}), 
whereas the 5-form term cancels with 
the imaginary part of the 3-form terms. 
Thus, we find a simple result 
\begin{equation}
\kappa \delta \psi_\mu = \partial_\mu \epsilon. 
\end{equation}
Therefore, $\delta\psi_\mu = 0$ requires that the transformation 
parameter $\epsilon$ is independent of $x^\mu$. 
\par
Substituting eqs.\ (\ref{killing}), (\ref{5form}), (\ref{3form3}) into 
eq.\ (\ref{susytrans}) the supertransformation 
of the Rarita-Schwinger field for $i=4,\cdots,9$ becomes 
\begin{eqnarray}
\left( {1+i\tau^* \over 1-i\tau}\right)^{-{1 \over 4}}
\kappa \delta \psi_i 
\A = \A \left( \partial_i +{1 \over 4}ie^\phi \partial_i\chi \right)
\tilde{\epsilon}
+\left( {1\over 4}\omega_i{}^{IJ}\Gamma_{IJ}
+{1\over 8}\partial_j \phi \hat{\Gamma}^j{}_i \right)
\tilde{\epsilon}
\nonu 
\A \A -{i\over 8}{x^j \over r} e^\phi \tilde{F}_{0123r} 
(hh')^{-1}f\Gamma^{0123}\hat{\Gamma}^j \hat{\Gamma}_i
\tilde{\epsilon}
\nonu
\A \A 
-{1\over 64} e^{2\phi} f \left(P_r\right)^{-1}{x^j\over r}
\left(\hat{\Gamma}_i{}^j - 3\delta_i^j\right)
\nonu \A \A \times
\left[ 2(hh')^{-1} \Gamma^{0123}{\cal G}_{01r}{\cal G}_{23r}
+ h'^{-2} {\cal G}_{01r}^2 - h^{-2} {\cal G}_{23r}^2 \right]
\tilde{\epsilon}. 
\end{eqnarray}
There are four kinds of gamma matrix structures on the 
right-hand side: $\Gamma_i{}^j$, 
$\Gamma_i{}^j \Gamma^{0123}$, $\delta_i^j$ 
and $\delta_i^j \Gamma^{0123}$. 
The terms with the first two structures are shown to vanish. 
The remaining terms are 
\begin{eqnarray}
\left({1+i\tau^* \over 1-i\tau}\right)^{-{1 \over 4}}
\kappa \delta \psi_i 
\A = \A 
\partial_i \tilde{\epsilon} 
+ {i\over 4}\sin \theta \sinh \theta' (hh')^{1/2}f^{-2}\partial_i f
\tilde{\epsilon} 
\nonu
\A + \A {3\over 16} (h\cos^2 \theta + h' \cosh^2 \theta')
f^{-1}\partial_i f \tilde{\epsilon} 
\nonu 
\A + \A 
{i\over 4} \cos \theta \cosh \theta' (hh')^{1/2} 
f^{-1} \partial_i f \Gamma^{0123} \tilde{\epsilon}. 
\end{eqnarray} 
The functions on the right-hand side can be written as total 
derivatives 
\begin{eqnarray}
{i\over 4} \sin \theta \sinh \theta' (hh')^{1/2} 
f^{-2}\partial_i f 
\A = \A -{1 \over 4} \partial_i 
\log {h'^{1/2}\sinh \theta' -i h^{1/2} \sin \theta 
\over h'^{1/2}\sinh \theta' +i h^{1/2} \sin \theta}, \nonu
{3\over 16} (h\cos^2 \theta + h' \cosh^2 \theta') 
f^{-1}\partial_i f
\A = \A -{3\over 16}\partial_i \log ( hh'f^{-2} ), \nonu
{i \over 4} \cos \theta \cosh \theta' (hh')^{1/2} 
f^{-1} \partial_i f 
\A = \A {i \over 4} \partial_i \log 
{h'^{1/2}\cosh \theta' + h^{1/2} \cos \theta 
\over h'^{1/2}\cosh \theta' - h^{1/2} \cos \theta}. 
\end{eqnarray}
It is easy to see that the condition $\delta\psi_i = 0$ determines 
the transformation parameter as 
\begin{eqnarray}
\epsilon\A =\A (hh'f^{-2})^{3 \over 16}
\left({1 + i \tau^* \over 1 - i \tau}\right)^{1 \over 4} 
\nonu 
\A\A \times 
\left[{h'^{1/2}\sinh \theta' - i h^{1/2} \sin \theta 
\over 
h^{1/2}\sinh \theta' + i h^{1/2} \sin \theta } \right]^{1 \over 4}
\left[{h'^{1/2}\cosh \theta' - i h^{1/2} \cos \theta \Gamma^{0123} 
\over 
h'^{1/2}\cosh \theta' + i h^{1/2} \cos \theta \Gamma^{0123}}
\right]^{1 \over 4}
\tilde{\epsilon}_0, 
\label{solution}
\end{eqnarray}
where $\tilde{\epsilon}_0$ is an arbitrary constant spinor. 
Substituting eq.\ (\ref{solution}) into the condition 
(\ref{killing2}) we find 
\begin{equation}
\tilde{\epsilon}_0^* = \Gamma^{01}\tilde{\epsilon}_0. 
\label{spinor}
\end{equation}
\par
Eq.\ (\ref{solution}) with a constant $\tilde\epsilon_0$ satisfying 
the condition (\ref{spinor}) is our final result on the parameter 
of unbroken supersymmetry. From eq.\ (\ref{spinor}) half of the 32 
supersymmetries are preserved. 
Thus, we have 16 supersymmetries in four dimensions for the 
solution with two non-vanishing components of the 2-form fields. 
Taking $\theta'=0$ it goes back to the solution for one 
non-vanishing component, and further taking the decoupling 
limit $r\rightarrow 0$ with some parameters kept constant, 
it becomes AdS${}_5$ $\times$ S${}^5$ in the near horizon 
limit \cite{MR}. These particular cases also have 16 
supersymmetries in four dimensions. 
\par
This result should be compared with the one in the case of 
vanishing $B_{MN}$ ($\theta = \theta' = 0$). 
In this case the $\delta\lambda = 0$ condition is automatically 
satisfied and one does not obtain eq.\ (\ref{killing}), 
which relates $\epsilon^*$ to $\epsilon$. 
The unbroken supersymmetries are 16 $x^\mu$-independent 
$\epsilon$ satisfying $i \Gamma^{0123} \epsilon = \epsilon$ 
and 16 $x^\mu$-dependent $\epsilon$ satisfying 
$i \Gamma^{0123} \epsilon = -\epsilon$ \cite{GP}. 
The former 16 correspond to Poincar\'e supersymmetry 
in the four-dimensional view point, while the latter 16 
correspond to conformal supersymmetry. 
\par
%
\vspace{10mm}
\noindent {\Large{\bf Acknowledgements}} 
\vspace{3mm}
\par
I would like to thank A. Tsuchiya for suggesting the topic
and K. Hashimoto for fruitful comments. 
I also would like to thank the Center for Theoretical Physics 
of MIT for hospitality. 
This work is supported in part by the Nishina 
Memorial Foundation and by funds provided by 
the U.S. Department of Energy (D.O.E.) under cooperative 
research agreement \#DF-FC02-94ER40818. 
%
%
\newcommand{\NP}[1]{{\it Nucl.\ Phys.\ }{\bf #1}}
\newcommand{\PL}[1]{{\it Phys.\ Lett.\ }{\bf #1}}
\newcommand{\CMP}[1]{{\it Commun.\ Math.\ Phys.\ }{\bf #1}}
\newcommand{\MPL}[1]{{\it Mod.\ Phys.\ Lett.\ }{\bf #1}}
\newcommand{\IJMP}[1]{{\it Int.\ J. Mod.\ Phys.\ }{\bf #1}}
\newcommand{\PR}[1]{{\it Phys.\ Rev.\ }{\bf #1}}
\newcommand{\PRL}[1]{{\it Phys.\ Rev.\ Lett.\ }{\bf #1}}
\newcommand{\PTP}[1]{{\it Prog.\ Theor.\ Phys.\ }{\bf #1}}
\newcommand{\PTPS}[1]{{\it Prog.\ Theor.\ Phys.\ Suppl.\ }{\bf #1}}
\newcommand{\AP}[1]{{\it Ann.\ Phys.\ }{\bf #1}}
\newcommand{\ATMP}[1]{{\it Adv.\ Theor.\ Math.\ Phys.\ }{\bf #1}}
\end{document}